# Assessing Bias in the Variable Bandpass Periodic Block Bootstrap Method

Yanan Sun[1*], Eric Rose[1], Kai Zhang[2], Edward Valachovic[1]

[1]Department of Epidemiology and Biostatistics, College of Integrated Health Science, University at Albany - State University of New York, Rensselaer, New York State, The United States of America

[2]Department of Population and Community Health, College of Public Health, The University of North Texas Health Science Center at Fort Worth, Fort Worth, Texas State, The United States of America

*Corresponding Author

Email: ysun7@albany.edu

# Abstract

The Variable Bandpass Periodic Block Bootstrap (VBPBB) is an innovative method for time series with periodically correlated (PC) components. This method applies bandpass filters to extract specific PC components from datasets, effectively eliminating unwanted interference such as noise. It then bootstraps the PC components, maintaining their correlation structure while resampling and enabling a clearer analysis of the estimation of the statistical properties of periodic patterns in time series data. While its efficiency has been demonstrated in environmental and epidemiological research, the theoretical properties of VBPBB—particularly regarding its bias of the estimated sampling distributions—remain unexamined. This study investigates issues regarding biases in VBPBB, including overall mean bias and pointwise mean bias, across a range of time series models of varying complexity, all of which exhibit periodic components. Using the R programming language, we simulate various PC time series and apply VBPBB to assess its bias under different conditions. Our findings provide key insights into the validity of VBPBB for periodic time series analysis and offer practical recommendations for its implementation, as well as directions for future theoretical advancements.

Key Words: Time Series, Periodically Correlated, Variable Bandpass Periodic Block Bootstrap, Seasonality, Pointwise Mean, Statistical Estimator, Sampling Distribution, Bias Adjustment

# Introduction

One primary objective of time series analysis is to determine the presence of periodic patterns, assess their recurrence, and evaluate the frequency at which they repeat over time.[1,2] Identifying such patterns is fundamental to understanding temporal regularity. Many real-world challenges involve time series data, and analyzing their periodic patterns can offer valuable insights. Examples include climate change, air quality fluctuations, and seasonal variations in disease mortality.[3-5] Furthermore, studying periodic patterns in user-product interactions can be highly beneficial for businesses, enabling them to refine marketing strategies, enhance user experience, and improve product recommendations.

To effectively analyze periodic variations in time series data, a novel model-free bandpass bootstrap approach, known as the Variable Bandpass Periodic Block Bootstrap (VBPBB), was introduced by Valachovic in 2024[6]. This method leverages Kolmogorov-Zurbenko Fourier Transforms (KZFT) to separate periodically correlated (PC) principal components from interfering frequencies and noise, followed by block bootstrapping of the extracted PC components[7,8]. VBPBB enables the investigation of periodic variations in time series data and facilitates the visualization and significance testing through confidence interval (CI) bands for the periodic means of the PC components. This capability makes VBPBB a powerful tool for uncovering hidden periodic trends in complex time series data, further strengthening its applications in many fields[4,5].

A key advantage of VBPBB is its ability to preserve the structural integrity of PC components after passing through KZFT, ensuring that the periodic characteristics remain intact during resampling. Moreover, an extension of VBPBB, the Variable Multiple Bandpass Periodic Block Bootstrap (VMBPBB) is designed to handle time series with multiple periodically correlated (MPC) components, where MPC refers to the presence of multiple PC components within a single time series[9]. Such structures commonly arise in real-world applications, where observed periodicity results from the cumulative influence of multiple underlying processes as well as the presence of harmonic fluctuation occurring at integer multiples of the fundamental frequency. By applying periodic block bootstrap to the extracted PC components, VBPBB enhances the estimation of periodic characteristics within both the individual component processes and the overall time series.

Efron in 1979 introduced a statistical method, which is called Bootstrap.[10] The bootstrap method is a powerful way of resampling, and it quickly gained widespread popularity, particularly in the context of model selection, where it became a valuable tool for evaluating and comparing different models based on their predictive performance and complexity. However, this method has been proven to have some problems, where the bias is one of the problems.[10-11] For instance, Young et al. in 1990 provided two simulated examples in their article, one with 10 observations and another with 20, and their simulation study demonstrated that the bootstrap method is noticeably biased when applied to small sample sizes in both cases.[11]

While bootstrap methods have been extensively studied, research specifically focused on block bootstrap techniques is relatively limited. Block bootstrapping is a variant of the traditional bootstrap, and it involves resampling contiguous blocks of data points rather than individual observations. This approach is particularly well-suited for time series data, where observations are typically correlated over time. Several types of block bootstrap methods have been developed. The Moving Block Bootstrap (MBB) was introduced by Künsch in 1989 and further extended by Liu and Singh in 1992.[12] A Seasonal Block Bootstrap, proposed by Politis, constrains block lengths to be integer multiples of a known period $p$.[13] In this approach, all resampled blocks begin and end at sequential positions within the cycle, regardless of the block's starting point. Other block bootstrap methods specifically designed for PC time series have been introduced, including the approaches of Chan et al. and the Generalized Seasonal Block Bootstrap (GSBB) proposed by Dudek et al.[14] These methods, collectively referred to here as Periodic Block Bootstraps (PBB), are particularly tailored to preserve periodic structures during resampling. Given the close relationship between PBB and standard bootstrap techniques, biases observed in classical bootstrap procedures may also arise in PBB, potentially affecting the accuracy of VBPBB results. Understanding and evaluating such bias is therefore essential when applying VBPBB to periodic time series data.

Since VBPBB is a newly developed method, its applications in various fields and potential limitations remain largely unexplored, with bias being a key issue that warrants careful consideration. VBPBB integrates multiple techniques to analyze periodic structures in time series data. The process begins with a periodogram, which is used to identify potential frequencies with higher power. A frequency exhibiting higher power suggests a significant deviation from other frequencies, indicating a possible periodic pattern within the time series. The periodogram, as described by Wei (1990), represents the spectral density or frequency domain characteristics of a time series.[15] There are various approaches to getting the frequencies or periods that are potentially significant, as well as visualizing the periodogram.

The KZFT bandpass filter, a repeated moving average technique, extracts periodic components while reducing noise but may over smooth the data, leading to biased estimates of PC components. Additionally, KZFT struggles to separate fundamental frequencies from harmonics due to its iterative smoothing. To address dependencies in the time series, we apply block bootstrap to the PC components, though improper block size selection and small sample size can distort temporal dependence and underestimate variability. These are the potential conditions that we believe may introduce bias but are not limited to those possible reasons; however, the actual impact must be validated through data analysis. A time series with periodic patterns can be regarded as a combination of multiple sine waves with various noise components. To systematically investigate bias, we generate multiple time series under different conditions, varying key parameters such as period, amplitude, phase, and noise levels. By considering scenarios ranging from simple to complex, we aim to examine potential biases in the PC components extracted from KZFT as well as any biases introduced by the block bootstrap procedure.

To evaluate the bias in the results obtained from the VBPBB method, we will assess the bias using two key measures: overall mean bias and pointwise mean bias. In statistical analyses, bias is commonly defined as the difference between the expected value of an estimator and the true parameter being estimated. When estimating a constant mean, the bias is straightforward, typically expressed as the deviation between the estimated mean and the true mean. However, in situations involving periodic or cyclical parameters, such as seasonal data, bias can vary significantly across different points in the cycle, making a simple overall measure insufficient.

The overall mean bias captures the average deviation of the estimated mean across the entire time series from the true mean, providing a general indication of whether the estimator systematically overestimates or underestimates the parameter globally. Conversely, pointwise mean bias assesses localized deviations at each specific time point, revealing how accurately the periodic structure is preserved throughout the cycle. By employing these two complementary measures, we can gain comprehensive insights into both global trends and local deviations, enhancing our understanding of the strengths and limitations of the VBPBB method in estimating periodic data.

# Methods

## Simulation Data Generation

To evaluate the performance of the VBPBB method under a range of signal and noise conditions, we conducted a simulation study using multiple synthetic time series. Each dataset was generated by combining sine waves with varying periods and amplitudes to mimic different types of periodic structure. Specifically, the true value of the signal for dataset $d$ was generated as:

$$y_t^{\text{true},(d)} = \sum_{j=1}^{J_d} A_j^{(d)} \cdot \sin\left(\frac{2\pi t}{P_j^{(d)}} + \phi_j^{(d)}\right), \quad t = 1, \ldots, T$$

where $J_d$ is the number of sine wave components in the $d$-th simulated dataset., $A_j^{(d)}$ is the amplitude in the $d$-th simulated dataset, $P_j^{(d)}$ is the period, and $\phi_j^{(d)}$ is the phase shift. $T$ in our study is 2500.

To simulate more realistic conditions, we added noise $\varepsilon_t^{(d)}$ to the periodic signals using different distributions. Thus, the observed value at time $t$ is:

$$y_t^{(d)} = y_t^{\text{true},(d)} + \varepsilon_t^{(d)}$$

## Period Detection and PC Component Extraction

For each simulated dataset, we first applied a periodogram to identify dominant frequencies, which were then converted to their corresponding period values. Both the frequencies and their associated periods were used as inputs to the KZFT filter. The KZFT was applied using a fixed iteration $k = 1$, and window size $m$ was chosen based on prior experience. The filter produced a noise-reduced signal containing the PC components.

In this simulation study, the true value at time $t$, $y_t^{\text{true},(d)}$, was known. However, for consistency with real-world settings, we used PC components extracted by the KZFT filter as pseudo-truths for bias estimation in some analyses. The formula is as follows:

$$y_t^{PC,(d)} = \sum_{i=1}^{K_d} y_t^{PC_i,(d)}$$

Where $y_t^{PC_i,(d)}$ is the value at time $t$ for $i$-th PC component for data set $d$, and $K_d$ is the number of PC components extracted from the KZFT bandpass filter for data set $d$.

## VBPBB Resampling and Estimation

The VBPBB procedure was applied to the PC components of each dataset. PBB resampling was performed using block sizes corresponding to the identified periods. For each dataset, $B = 1000$ bootstrap samples. The resulting resampled matrix $B \times T$ was used to estimate CI and various types of bias.

## Overall Mean and Overall Mean Bias

The overall mean serves as one benchmark for bias assessment in this study. It represents the average value of the entire time series and is used to evaluate potential deviations in different steps of the analysis. The overall mean of the original time series is calculated as the arithmetic mean of all observations. The overall mean captures the global average level of the series, ignoring any periodic or time-dependent structure. In this simulation study, the true overall mean is analytically defined based on the sine wave used to generate data. The corresponding formula is as follows:

$$\theta^{(d)} = \frac{1}{T} \sum_{t=1}^{T} y_t^{\text{true},(d)}$$

Since the true overall mean is not accessible in real-word settings, we use the overall mean of the extracted PC components as a practical alternative if necessary in future research.

To better evaluate the bias introduced by the periodic block bootstrap procedure, we use the mean of the sum of the extracted PC components as the sample mean. The formula is as follows:

$$\bar{y}^d = \frac{1}{T} \cdot y_t^{PC,(d)}$$

The overall mean of the VBPBB results is defined as follows:

$$\hat{y}^{*(d)} = \frac{1}{T \cdot B} \sum_{t=1}^{T} \sum_{b=1}^{B} y_t^{*(b)(d)}$$

Therefore, the overall mean bias is defined by:

$$\text{Bias}_{\text{overall}} = \hat{y}^{*(d)} - \theta^{(d)}$$

In real-world applications where $\theta^{(d)}$ is not available, we estimate the bias using the sample mean of the sum of PC components:

$$\widehat{\text{Bias}}_{\text{overall}} = \hat{y}^{*(d)} - \bar{y}^d$$

### Pointwise Mean Bias and Periodic Mean Bias

To assess the performance of the VBPBB method at each time point, we compute the pointwise mean across multiple bootstrap replicates. Specifically, for each time $t = 1, \ldots, T$, the pointwise mean is defined as the mean of the corresponding estimates obtained from $B$ independent bootstrap samples, and we define the pointwise mean for the $d$-th simulation as follows:

$$\hat{y}_t^{*(d)} = \frac{1}{B} \sum_{b=1}^{B} y_t^{*(b)(d)}$$

Therefore, the pointwise mean bias is defined as follows:

$$\text{Bias}_{\text{pointwise}} = \hat{y}_t^{*(d)} - \theta_t^{(d)}$$

Where $\theta_t^{(d)} = y_t^{\text{true},(d)}$ denotes the true value at time $t$ in the $d$-th simulation.

In real-world applications where $\theta_t^{(d)}$ is not observable, we instead use the sample mean of the extracted principal component as a proxy:

$$\widehat{\text{Bias}}_{\text{pointwise}} = \hat{y}^{*(d)} - y_t^{\text{PC},(d)}$$

Typically, the pointwise mean is calculated at each individual time point. However, in practice, it is often meaningful to consider mean over defined cycles or periods. Thus, we introduce a special type of pointwise mean, called the periodic mean, which aggregates estimates across equivalent positions within each cycle. Formally, we define the periodic mean as the mean of estimated values grouped by periodic positions within a defined cycle of length $P$. $P^{(d)}$ is the period for $d$-th dataset. The periodic true mean at position $k$ is computed as follows:

$$\theta_k^{(d)} = \frac{1}{N_k} \sum_{t=1}^{T} \mathbf{1}_{\{t \equiv k \ (\text{mod} \ P^{(d)})\}} \cdot y_t^{\text{true},(d)}$$

For real-world data, the sample mean can be used instead:

$$\overline{y(t)}^d = \frac{1}{N_k} \sum_{t=1}^{T} \mathbf{1}_{\{t \equiv k \ (\text{mod} \ P^{(d)})\}} \cdot y_t^{PC,(d)}$$

Where each $k$ corresponds to a specific position within the cycle. For instance, if $k = 1$, then the formula above will compute the periodic mean of the first position for all periodic cycles. The range of $k$ depends on the period, $k \in \{1, \ldots, P^{(d)}\}$, which means if the period is $P^{(d)}$, there will be $P^{(d)}$ groups of time points with same position in different cycles. $N_k$ is number of time points that belong to phase k within the full time series of length $T$. Also, same as overall mean and pointwise mean, the true mean of periodic mean is computed based on the sine wave or sum of sine waves of $d$-th simulation.

The periodic mean of VBPBB is defined as follows:

$$\widehat{y(k)}^{*(d)} = \frac{1}{N_k} \sum_{t=1}^{T} \sum_{b=1}^{B} \mathbf{1}_{\{t \equiv k \ (\mathrm{mod} \ P^{(d)})\}} \cdot y_t^{*(b)(d)}$$

Where $\mathbf{1}_{\{t \equiv k \ (\mathrm{mod} \ P^{(d)})\}}$ is the indicator function, equal to 1 when time points $t$ belong to phase $k$, and 0 elsewhere.

The periodic mean bias is defined by:

$$\mathrm{Bias}_{\mathrm{periodic}} = \widehat{y(k)}^{*(d)} - \theta_k^{(d)}$$

In real-world applications where $\theta_k^{(d)}$ is not available, we estimate the bias using the sample mean of the sum of PC components:

$$\widehat{\mathrm{Bias}}_{\mathrm{periodic}} = \widehat{y(k)}^{*(d)} - \overline{y(t)}^d$$

In real-world applications, periodic means are often more relevant, and our study introduces and recommends the use of periodic mean bias as a new and practical metric. When the number of time points is small, pointwise bias remains flexible and informative. However, as time series become longer and the focus shifts toward capturing periodic structure, we recommend using periodic mean bias, which offers a more targeted and interpretable evaluation of periodic pattern estimation.

**Bias Correction**

In this study, we consider two types of bias. The first is defined concerning the true mean, which provides an ideal benchmark but is generally unavailable in real-world data. The second is defined using the sample mean, which is a practical substitute when the true mean cannot be determined. As discussed earlier, in most real-world applications the true mean is unknown; therefore, we additionally report bias estimates based on the sample mean to facilitate comparison and to better reflect realistic scenarios.

For bias correction, different strategies may be applied depending on the type of bias considered. When using the overall mean bias, a global adjustment can be made by directly shifting the estimated result based on the average bias across the entire series. In cases where the bias varies over time, such as when using pointwise mean bias, a localized correction may be necessary. This involves adjusting individual time points based on their specific estimated bias values, potentially leading to more accurate and nuanced correction of the estimated signal. For the periodic mean bias, we apply a localized correction by adjusting the bias for each individual period block.

## Simulation

We conducted a simulation study to assess the performance and robustness of our proposed methods. A synthetic time series of length was generated by sine wave with specified periods, amplitudes, and phase shifts or phase angles. The period chosen was 25 corresponding to hypothetical short-term patterns. The amplitudes were set to 0.8, with an additional phase angle 100. Formally, the signal, $y_t$, was generated as follows:

$$y_t = 0.8 \sin\left(\frac{2\pi}{25} t + 100\right)$$

To approximate real-world data conditions, we added noise to this clean signal. Specifically, we generated mixed noise comprising two distributions: normal and gamma, with mixture weights of 0.5 and 0.5, respectively. The noise parameters were set as follows: normal noise with a standard deviation of 3 and gamma noise with shape parameter 2 and scale parameter 5. To ensure the gamma noise had a negative mean and strong right skew, we subtracted 10 from the generated gamma values. Thus, the final observed time series $y^{noise}$ was generated by:

$$y^{noise} = y_t + 0.5 \cdot \text{Normal}(0, 9) + 0.5 \cdot (\text{Gamma}(2, 5) - 10)$$

An example realization of the simulated data and the true underlying signal is illustrated in Figure 1, clearly showing how the mixed noise affects the underlying periodic structure.

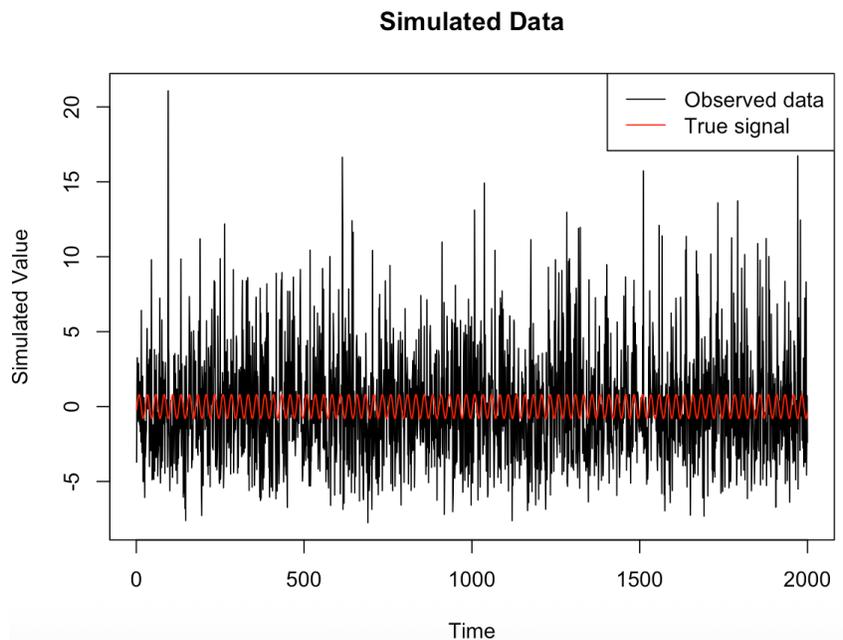

To simulate a sudden event such as a lockdown or pandemic, we artificially decreased the signal by 4 units between days 1500 and 1600. This created a sharp drop in the time series during this interval, mimicking the effect of a major intervention or disruption.

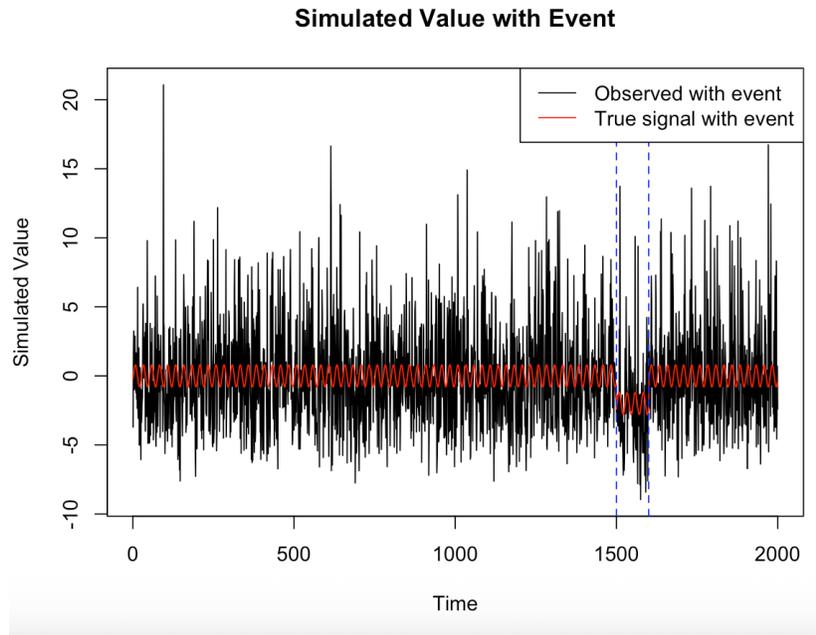

To further approximate real-world scenarios where time series values may gradually increase or decrease over time, for example, the PM2.5 values, we added a linear trend to the simulated data. Specifically, a linearly increasing trend was added to both the true underlying signal and the observed time series after noise was introduced. The trend was defined as a constant increment added at each time point, resulting in a gradual upward shift across the entire time series. In this study, the trend coefficient was set to 0.001 per time unit.

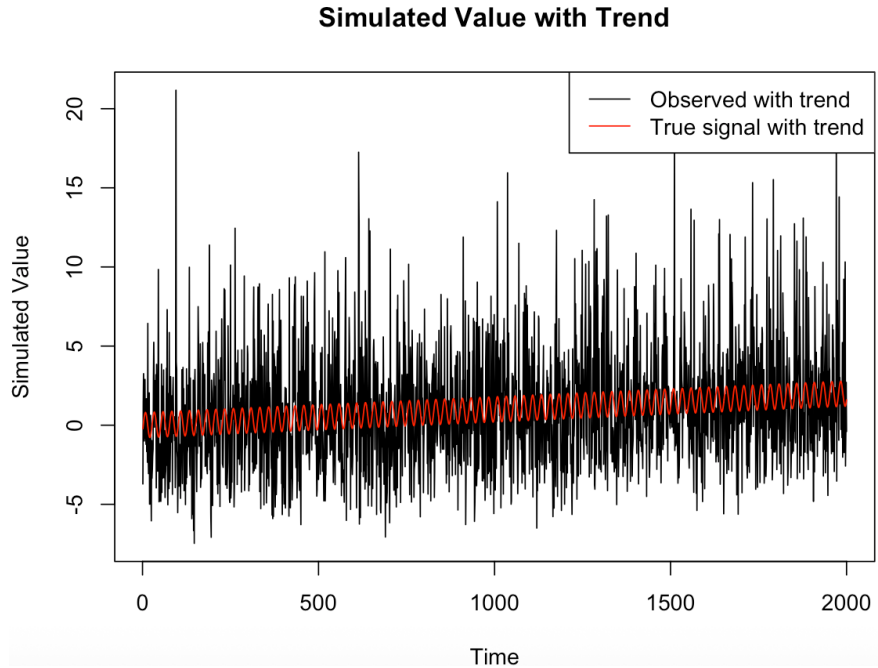

Simulated Value with Trend

# Results

We evaluated the performance of the proposed VBPBB method using simulated data with known ground truth. The method's effectiveness was assessed using the three complementary metrics: overall mean bias, pointwise mean bias, and periodic mean bias.

In this study, we applied the KZFT filter with the following parameters: $P = 25$, window size $m = 251$ (calculated as $25 \times 10 + 1$), number of iterations $k = 1$, and frequency $f = 1/25$. These settings were used for all three scenarios in the following results subsections. These settings were used consistently across all three scenarios described in the following results subsections. The window size m was selected to be a multiple of the period of interest ($P$) and was rounded to the nearest odd integer to satisfy the filter's requirement.

## Results Visualization

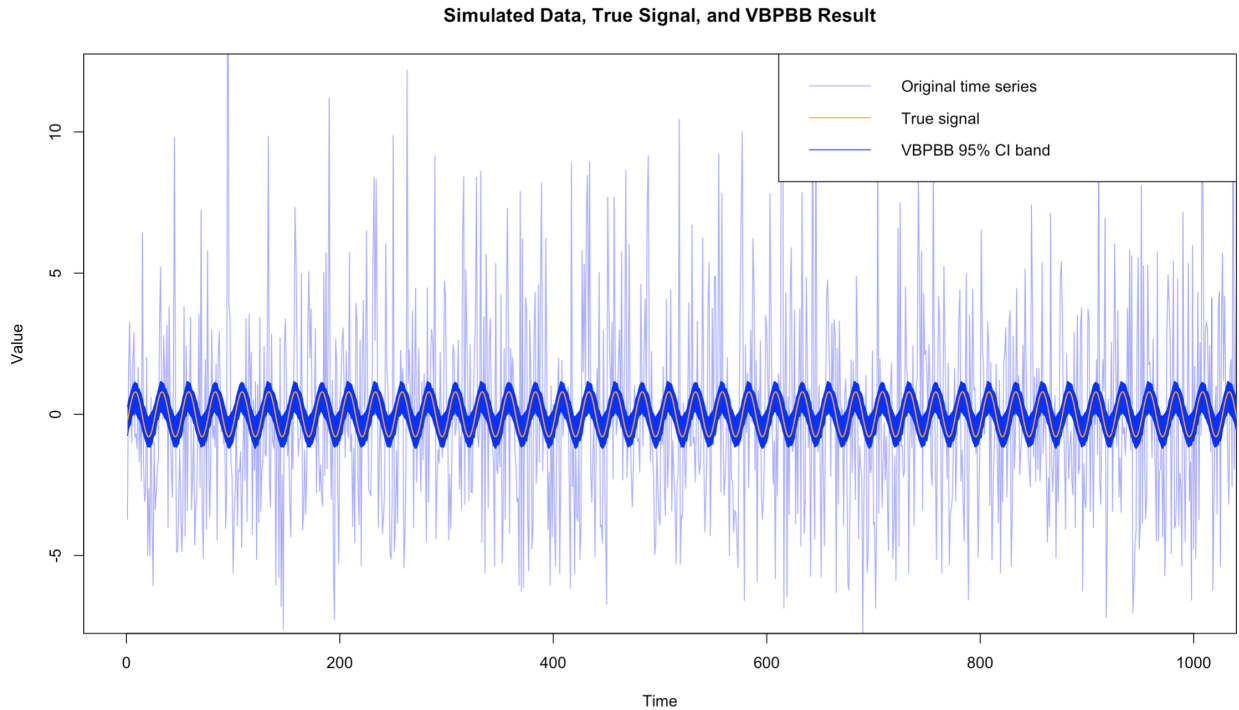

Figure 4. VBPBB results for the original simulated time series

Figure 4 presents the VBPBB results for the original simulated time series, and to better visualize the pattern, we display only the first 1,000 values of the series. Both visually and computationally, we observe that the orange-colored true signal remains entirely within the 95% CI band produced by VBPBB. This indicates that the method successfully captures the underlying periodic structure despite the presence of substantial noise.

It is important to note that this result is closely related to the choice of the smoothing argument $m$. When a larger m is used, the 95% CI band becomes narrower, which may result in the true signal falling outside the band. Therefore, the selection of m plays a critical role in balancing the trade-off between smoothness and coverage in periodic signal estimation.

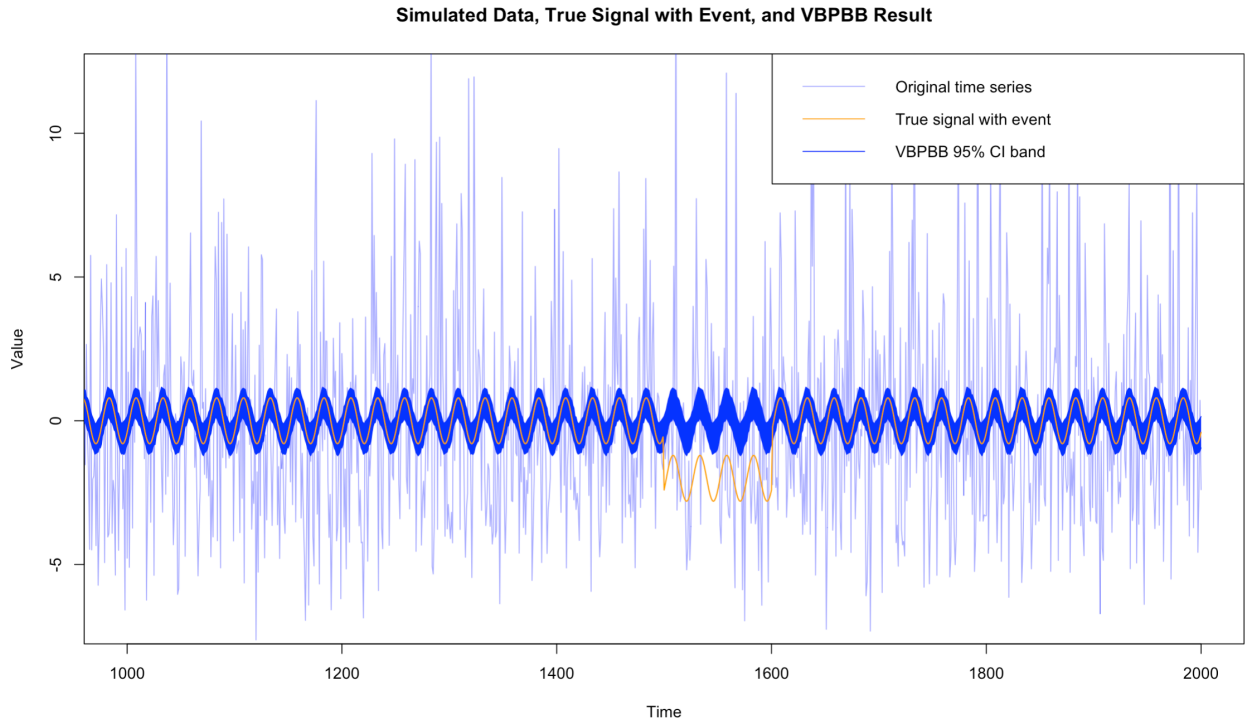

Figure 5. VBPBB results for the simulated time series with event

Figure 5 displays the VBPBB results for the simulated time series with an event, focusing on time points 1000 to 2000. This time window was selected because the event was introduced in this interval. As shown in the plot, there is a sudden downward shift in the true signal between time points 1500 and 1600, representing the event.

Outside of this short event period, the true signal (orange curve) remains well within the 95% CI band generated by VBPBB. However, during the event itself, the estimated band does not fully capture the drop. This is expected, as the VBPBB method provides a CI band only for the PC component of the signal. This illustrates a key characteristic of VBPBB: while it is effective at quantifying uncertainty in periodic structure, it is not designed to capture localized deviations that fall outside the modeled periodic pattern.

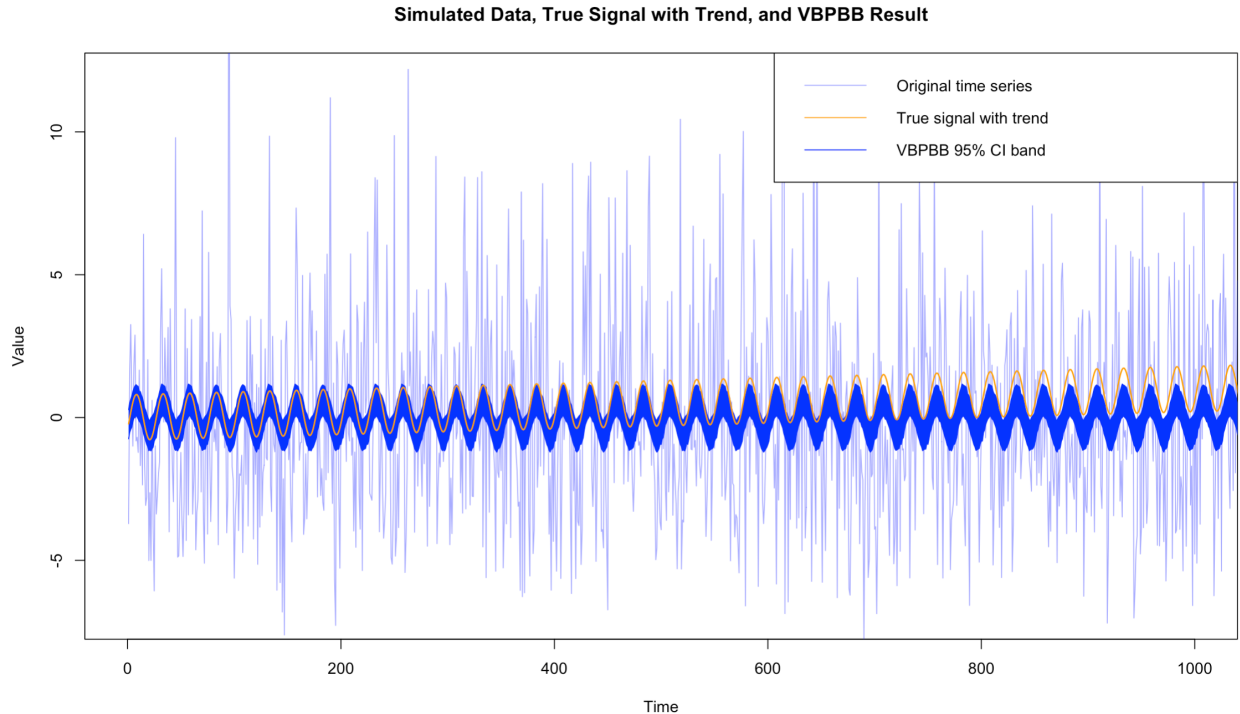

Figure 6. VBPBB results for the simulated time series with trend

Figure 6 shows the VBPBB results for the simulated time series with a linear trend, using the first 1,000 values for visualization. In this scenario, a gradual upward trend was added to the original periodic signal. As expected, the VBPBB 95% CI band does not follow this upward trend, since the method is specifically designed to extract periodic components and is not intended to capture non-periodic trends.

Although the VBPBB output may differ slightly from the original (non-trend) case due to variation in the extracted principal components, these differences are not visually prominent in the plot. Importantly, if we subtract the added trend from the true signal, the VBPBB result (obtained from the data with trend) still effectively captures the underlying periodic signal. This reinforces the robustness of VBPBB in isolating periodic patterns, even when the observed data include non-periodic structural components.

## Overall Mean Bias

The overall mean bias quantifies global deviation between the bootstrap-estimated signal and the true signal mean. Across simulations, the bias remained small in magnitude, indicating that the VBPBB method preserved the global level of the signal despite the presence of heavy-tailed and skewed noise, a linear trend, and a sudden drop. Since the true mean is known in the simulation study, the true overall mean bias was calculated using the true overall mean. The estimated overall mean bias was calculated using the sample overall mean.

Table 1.

| Simulation | True Overall Mean | Sample Overall Mean | VBPBB Overall Mean | True Overall Mean Bias | Estimated Overall Mean Bias |
|---|---|---|---|---|---|
| Original Simulated Value | 0 | -0.000483 | -0.000505 | -0.000505 | -0.000022 |
| Simulated Value with Event | 0 | -0.001288 | -0.001340 | -0.001340 | -0.000052 |
| Simulated Value with Trend | 0 | 0.007310 | 0.007299 | 0.007299 | -0.000011 |

Although the simulated value with event and simulated value with trend were modified versions of the original time series, we set the true overall mean to zero for all three scenarios. This is because the original simulated time series was constructed from a sine wave without vertical shifting, resulting in a theoretical overall mean close to zero. In general, introducing a downward shift over a specific segment (as in the event scenario) or adding a linear trend would change the global mean of the signal. However, the VBPBB method focuses on capturing periodic components by selecting PC components and resampling them. As such, it emphasizes the periodic structure of the signal rather than the global level. Since the goal of this study is to assess the bias of the VBPBB method in recovering periodic behavior, we standardized the true overall mean to zero across all scenarios to isolate and evaluate the effects of structural modifications (e.g., event or trend) on estimation bias without confounding from changes in the mean level.

As shown in Table 1, the overall mean bias across all simulation scenarios remained relatively close to zero, indicating that the VBPBB method preserved the global signal level to a reasonable extent. In this study, we do not attempt to judge whether these biases are substantively large or small. Instead, these values serve as reference points for potential future correction: if adjustment is needed in subsequent applications, the bias observed under similar conditions can be directly applied to align the VBPBB estimates with the original data scale.

**Periodic Mean Bias**

To further investigate how the estimation bias varies across different scenarios, we examined the periodic mean bias at each individual period. The following figure presents the bias values under the Original, Event, and Trend scenarios, allowing for a more detailed comparison of their behavior over time.

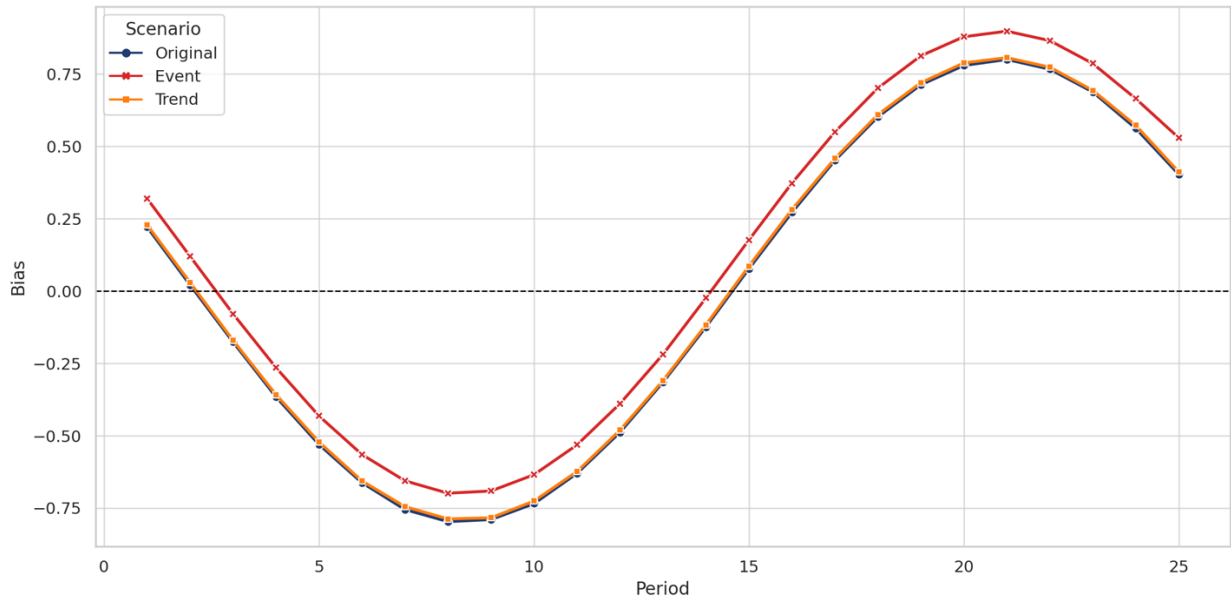

Figure 7. Bias per period based on the true periodic mean for the original simulated value, the simulated value with event, and the simulated value with trend.

Figure 7 illustrates the bias in estimated periodic means across 25 periods under three different scenarios: Original (blue), Event (red), and Trend (orange). The red curve shows a larger bias in several periods due to the presence of a localized disruption. In contrast, the Original and Trend scenarios produce nearly identical bias values across all periods, resulting in their respective curves (blue and green) appearing visually overlapping in many sections of the plot. This apparent overlap reflects the small numerical differences between the two scenarios rather than an absence of distinction.

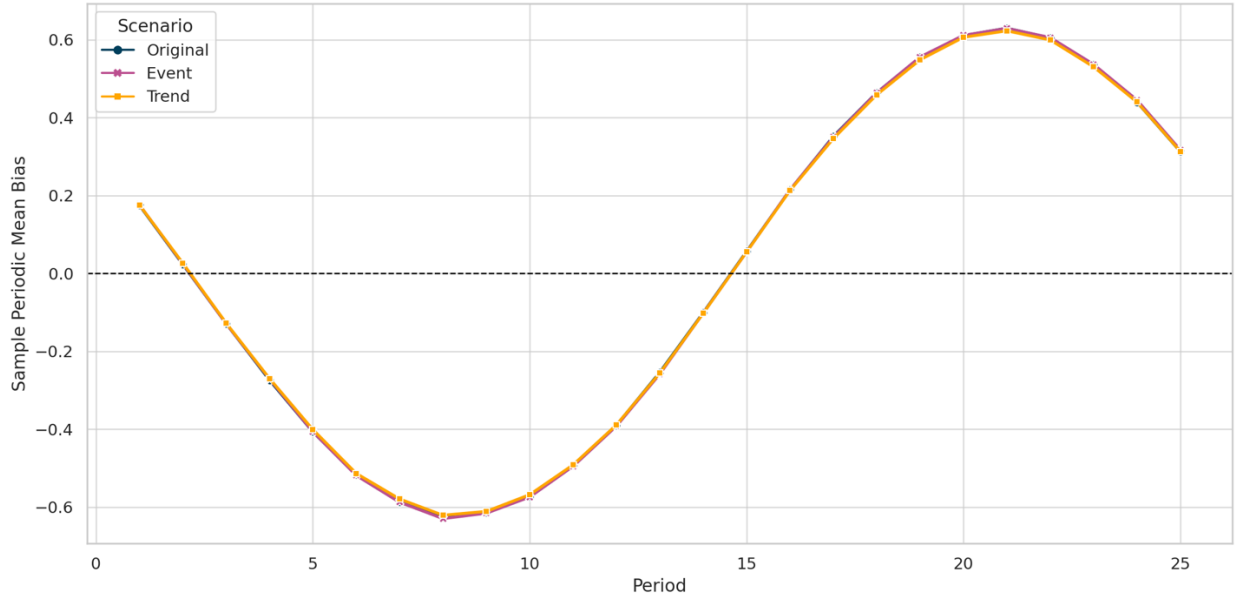

Figure 8. Bias per period based on the sample periodic mean for the original simulated value, the simulated value with event, and the simulated value with trend.

Figure 8 shows the bias calculated based on the sample periodic mean. As mentioned earlier, the sample mean in this study is computed on the PC component extracted from the data. In real-world applications, the sample mean will also use the result of the PC components. Since the PBB step is applied to the result of the PC component, which we obtain using the KZFT filter, it is important to note that KZFT is essentially a type of moving average filter. As such, it tends to smooth out the data. Therefore, any strong localized changes in the original signal may not be apparent in the filtered PC component. On the other hand, a stable trend will not affect the KZFT result either, since the output represents variation around the periodic pattern. This explains why the three bias curves in the figure (Original, Event, and Trend) are almost overlapping. The y-values of these curves represent the bias values that should be used for correction.

Since the VBPBB method is specifically designed to capture periodic variation, we focus our evaluation on the periodic mean bias. Rather than assessing deviation at every individual time point, the periodic mean bias aggregates the bias over predefined periodic intervals. This serves as a useful reference for potential bias correction in future applications.

For the true periodic means, we applied the same approach as described earlier for the overall mean. Specifically, we used the original simulated signal to compute the true periodic means and applied these same values across all scenarios, including the simulation with event and the simulation with trend. This decision is consistent with our previous statement for the overall mean: when calculating the true overall mean of the original simulated signal, the result was less than 0.0001, and thus we set it to zero. Following the same logic, we assigned a true overall mean of zero to the other two scenarios as well. In the periodic case, since we computed the mean for each periodic segment individually, the true periodic means vary by period. For consistency and to isolate the effect of structural changes, we adopted the original periodic means as the true values across all three simulation settings.

## Discussion

The KZFT bandpass filter is a type of moving average filter that emphasizes periodic variation, it allows us to extract periodically correlated components. The subsequent application of PBB preserves the correlation structure of these components during resampling, thereby enabling clearer analysis of the statistical properties of periodic patterns in time series data, where this is feature that has been well demonstrated. In the analysis of overall mean bias, we observed that bias tends to appear in the latter two scenarios: the simulated value with an event and the simulated value with a trend. These situations are commonly encountered in real-world data. While overall mean bias is a commonly used metric, this study introduces two additional methods: pointwise mean bias and the derived periodic mean bias. These approaches offer a more localized view of estimation bias and are not limited to the context of this study. In fact, they can be broadly applied to any project that involves observing, computing, or evaluating periodic changes. For example, in long-term observational data where the goal is to compare periodic means across weekdays (e.g., Monday through Sunday), periodic mean bias can serve as a useful diagnostic tool. Even when other time series methods are employed, this framework can provide a valuable reference for assessing potential bias and guiding correction strategies.

If greater alignment between the VBPBB output and the original data is desired—particularly in the presence of distinct events such as a prolonged shutdown—one potential strategy is to compute the mean of the original data during the event period and compare it to the corresponding mean from the VBPBB output. This comparison can then inform targeted adjustments to the VBPBB estimates for that specific segment, improving interpretability without altering the periodic structure preserved by the method. It is important to note that because the VBPBB method focuses on extracting periodic variation, its output is typically centered around zero. However, in many real-world datasets, such as PM2.5 measurements, the true overall mean can be substantially higher. In such cases, after estimating the periodic structure using VBPBB, the overall mean of the original data can be added back to the result to facilitate interpretation and ensure consistency with the observed data scale. However, if the focus is solely on observing long-term periodic variation, such adjustments may not be necessary.

As shown in the Results section, VBPBB does not aim to perfectly capture events or trends, as its primary goal is to reveal periodic variation and assess the significance of specific PC components. For example, in the context of PM2.5 data, concentrations may drop during specific events such as COVID-19 lockdowns, or exhibit a long-term downward trend due to improved air quality policies. In such cases, if the research objective is to present finer details in the estimated signal, adjustments can be made post-VBPBB processing. Specifically, one could first examine the data for potential events, compute the mean of the affected period and compare it to the rest of the series, and then adjust the VBPBB output accordingly when plotting. Similarly, if a long-term trend is present, it can be estimated in advance and added back to the VBPBB output during visualization to enhance interpretability. These adjustments can help better highlight specific data features that are not captured by the periodic-focused design of VBPBB. However, if the primary objective is to examine periodic variation, such as identifying a 7-day cycle in PM2.5 levels or testing the significance of a PC component, then the above corrections are not necessary.

As noted in the Results section, the primary goal of this study was not to evaluate whether the observed biases were large or small. Instead, our focus was on demonstrating several approaches to identify and quantify bias in the context of periodic signal estimation. These methods provide a foundation for addressing potential systematic bias introduced by bootstrap procedures. In future research, the bias metrics presented here can serve as reference points to guide appropriate bias correction strategies, thereby improving the reliability of bootstrap-based inference in time series analysis.

# References

1. Shumway RH, Stoffer DS. Time series analysis and its applications: with R examples. New York, NY: Springer New York; 2006 May.
2. Box GE, Jenkins GM, Reinsel GC, Ljung GM. Time series analysis: forecasting and control. John Wiley & Sons; 2015 May 29.

3. Mudelsee M. Climate time series analysis. Atmospheric and Oceanographic Sciences Library. 2010;397.

4. Sun Y, Valachovic EL. Seasonal and periodic patterns of PM2. 5 in Manhattan using the variable bandpass periodic block bootstrap. PloS one. 2025 Jun 25;20(6):e0326767.

5. Valachovic EL, Shishova E. Seasonal and periodic patterns in US COVID-19 mortality using the Variable Bandpass Periodic Block Bootstrap. PLoS One. 2025 Jan 22;20(1):e0317897.

6. Valachovic EL. Periodically correlated time series and the Variable Bandpass Periodic Block Bootstrap. PLoS One. 2024 Sep 17;19(9):e0310563.

7. Zurbenko I. The spectral analysis of time series. Amsterdam: North-Holland 1986.

8. Yang W, Zurbenko I. Kolmogorov–zurbenko filters. Wiley Interdisciplinary Reviews: Computational Statistics. 2010 May;2(3):340-51.

9. Valachovic E. The Variable Multiple Bandpass Periodic Block Bootstrap for Time Series With Multiple Periodic Correlations. Journal of Probability and Statistics. 2025;2025(1):9968540.

10. Efron B. Bootstrap Methods: Another Look at the Jackknife. The Annals of statistics. 1979;7(1):1-26.

11. Young GA, Daniels HE. Bootstrap bias. Biometrika. 1990 Mar 1;77(1):179-85

12. Liu RY, Singh K. Moving blocks jackknife and bootstrap capture weak dependence. Exploring the limits of bootstrap. 1992;225:248.

13. Politis DN. Resampling time series with seasonal components. in Frontiers in data mining and bioinformatics: Proceedings of the 33rd symposium on the interface of computing science and statistics:13–17 2001.

14. Dudek AE, Leskow J, Paparoditis E et al., A generalized block bootstrap for seasonal time series. Journal of Time Series Analysis. 2014;35(2):89–114.
15. Wei W. Time series analysis: univariate and multivariate methods. Redwood City, Calif: Addison-Wesley Pub. 1990.